\documentclass[12pt]{iopart}
\usepackage{epsfig}

\newcommand{\fig}[1]{Fig.~\ref{#1}}
 
\begin{document}

\title{Improving a radiative plus collisional energy loss model for application to RHIC and LHC}
\vspace{-0.2in}
\author{Simon Wicks, Miklos Gyulassy}
\address{Columbia University, Dept of Physics, 538 West 120th Street, New York, NY 10027}
\ead{simonw@phys.columbia.edu}
\vspace{-0.2in}
\begin{abstract}
With the QGP opacity computed perturbatively and with the global entropy constraints imposed by the observed $dN_{ch}/dy\approx1000$, radiative energy loss alone cannot account for the observed suppression of single non-photonic electrons. Collisional energy loss is comparable in magnitude to radiative loss for both light and heavy jets. Two aspects that significantly affect the collisional energy loss are examined: the role of fluctuations, and the effect of introducing a running QCD coupling as opposed to the fixed $\alpha_s=0.3$ used previously.
\end{abstract}

\vspace{-0.5in}
\section{Introduction}
\vspace{-0.2in}
Non-photonic single electron data~\cite{elec:PHENIX,elec:STAR}, which present an indirect probe of heavy quark energy loss, have
significantly challenged the underlying assumptions of jet tomography theory.  A much larger suppression of electrons than predicted~\cite{Djordjevic:2005db} was observed in the $p_T\sim 4-8 $ GeV region. ``These data falsify the assumption that heavy quark quenching is dominated by [pQCD based] radiative energy loss when the bulk [weakly coupled] QCD matter parton density is constrained by the observed dN/dy $\approx$ 1000 rapidity density of produced hadrons.'' \cite{Wicks:2005gt}

WHDG \cite{Wicks:2005gt} revisited the assumption that pQCD collisional energy loss is negligible compared to radiative energy loss~\cite{Gyulassy:2000er,Djordjevic:2003zk}. As argued there, and references therein, ``the elastic component of the energy loss cannot be neglected when considering pQCD jet quenching.'' As shown in WHDG and elsewhere \cite{Renk:2006pk}, the computationally expensive integrations over the geometry of the QGP cannot be reduced to a simple `average length' prescription. Indeed, this computation time is essential to produce radiative + collisional energy loss calculations consistent with the pion data.

There are large theoretical uncertainties in the WHDG results~\cite{Wicks:2007am}. Very significant to the electron prediction is the uncertainty in the charm and bottom cross-sections. There are also theoretical uncertainties in the energy loss mechanisms. Here, two aspects of the collisional energy loss will be examined with the aim of improving the energy loss model.
\vspace{-0.25in}
\section{Collisional energy loss fluctuations}
\vspace{-0.2in}
Similar to radiative energy loss, the fluctuations of collisional energy loss around the mean affect the quenching of the quark spectra. Collisional fluctuations are often modelled in a Fokker-Planck formalism, characterized by two numbers or functions: drag and diffusion. WHDG implemented an approximation to this scheme applicable for small energy loss by giving the collisional loss a gaussian width around the mean, with $\sigma^2 = 2 T \langle\epsilon\rangle$, where $\langle\epsilon\rangle$ is the mean energy loss given by a leading log calculation.

The drag-diffusion method is essentially a continuum approximation to a discrete process. A high energy jet traversing the QGP will undergo only a small number of collisions. In the Gyulassy-Wang model, the expected mean free path of a quark is $\sim2$fm, so there is a very significant surface region in which the fluctuations will differ greatly from those given by the continuum approximation. It is therefore necessary to look at the fluctuations per collision and in the number of collisions. A simple model to investigate this is to model the medium as \textit{initially} static objects which will then recoil upon collision, model the interaction between jet and medium using the full HTL medium modified propagator. This gives the probability of longitudinal momentum loss:
\begin{eqnarray}
  \frac{dN}{dq_{\Vert}} = \frac{\rho L g^4 }{4 \pi}\frac{C_R C_2}{d_A} \frac{m E}{E+m} \left[ C_L |\Delta_L|^2 + C_T |\Delta_T|^2 \right]  \nonumber \\
C_L = 2+\frac{1}{E}(\omega + \vec{v}.\vec{q})(2 - \frac{\omega}{m})\,,\, C_T = \left( \frac{-\omega}{m}\right)\left( v^2 - (\vec{v}.\hat{\vec{q}})^2 \right)
\end{eqnarray}

This single collision distribution is then Poisson convoluted to give the distribution for a finite number of expected collisions:
\begin{equation}
 P_{n+1}(\epsilon,E) \frac{e^{-\langle N_{coll}(E) \rangle}}{(n+1)!} \int dx_1 \ldots dx_n \rho(x_1,E) \ldots \rho(x_n,E) \rho(\epsilon - x_1 - \ldots - x_n,E)
\end{equation}

\begin{figure}[!th]
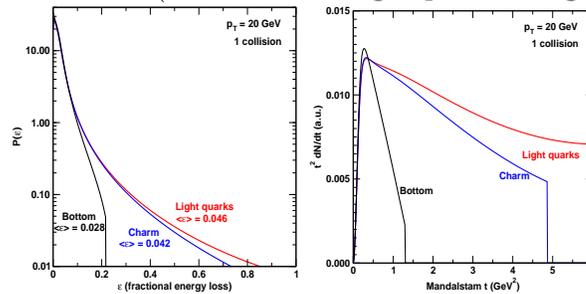
 
\centering
\epsfig{file=allflavs_n1_pofeps.eps,height=1.5in,width=1.5in,clip=,angle=0} 
\epsfig{file=allflavs_n1_t2dNdt.eps,height=1.5in,width=1.5in,clip=,angle=0} 
\caption{ \noindent
\label{fig:pofeps}
The distribution of fractional energy loss $\epsilon$ (left) and Mandalstam variable $t$ (right - scaled by $t^2$) in a collision using this model.
\vspace{-0.2in}
}
\end{figure}

The mass of the medium particle is tuned to give an average energy loss similar to that of the BT and TG leading log calculations ($m\sim0.2$GeV - although here we are not interested in the average energy loss per se). In \fig{fig:pofeps}, the probabiliy of fractional energy loss in one collision is shown, similar to a $1/t^2$ Bjorken collisional style model, with screening at small t-values (shown in the right pane of \fig{fig:pofeps}).
\vspace{-0.2in}

\begin{figure}[!th]
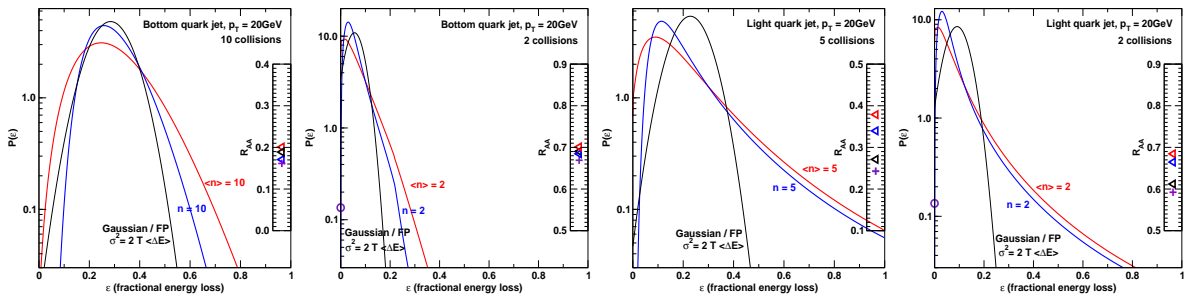
 
\centering
\epsfig{file=bottom_n10_pofeps_v2.eps,height=1.5in,width=1.5in,clip=,angle=0} 
\epsfig{file=bottom_n2_pofeps_v2.eps,height=1.5in,width=1.5in,clip=,angle=0} 
\epsfig{file=light_n5_pofeps_v2.eps,height=1.5in,width=1.5in,clip=,angle=0} 
\epsfig{file=light_n2_pofeps_v2.eps,height=1.5in,width=1.5in,clip=,angle=0} 
\caption{ 
\label{fig:collFluctBtm} 
The distribution in fractional energy loss for a 20 GeV jet, for a bottom quark jet and a light quark jet, for different numbers of collisions. Shown are lines for the gaussian approximation, exactly n collisions (labelled eg `n=10') and on average n collisions (labelled eg `$\langle n \rangle =10$'). Inset is the $R_{AA}$ for these distributions and the $R_{AA}$ evaluated for a delta function distribution at the mean loss (+ point).
\vspace{-0.2in}
\label{fig:collFluct} 
}
\end{figure}

Figure~\ref{fig:collFluct} illustrates the distributions in energy loss for a finite number of collisions for bottom and light quark jets. The results for charm quarks are qualitatively similar to those for light quarks. For a large number of collisions (eg average number of collisions $\langle n \rangle = 10$, L$\sim20$fm), the distributions are roughly symmetric and somewhat similar to the simple WHDG gaussian. This is expected from the central limit theorem. The $R_{AA}$ values extracted from these distributions are similar, with $\langle n\rangle=10$ and the gaussian approximation only differing by $\sim0.01$. Surprisingly, a similar result for the $R_{AA}$ values is found for $\langle n\rangle=2$ collisions for bottom quarks.
The large change arrives for light quarks. For both $\langle n \rangle =2,5$ collisions, the gaussian approximation gives a very different distribution for the fluctuations and a very different $R_{AA}$ value. The gaussian approximation overpredicts the $R_{AA}$ suppression by $~0.1$, which is around a 30\% effect for $\langle n \rangle=2$ collisions. This cannot be neglected. A full treatment of the finite number of collisions will reduce the quenching due to elastic energy loss compared to the treatment in WHDG. This conclusion is also applicable to other uses of Fokker-Planck / Langevin formalisms that use a continuum description of the collisional process. The $R_{AA}$ predictions for bottom quarks are likely only marginally affected, those for light quarks most affected.
\vspace{-0.25in}
\section{Running QCD coupling}
\begin{figure}[!th]
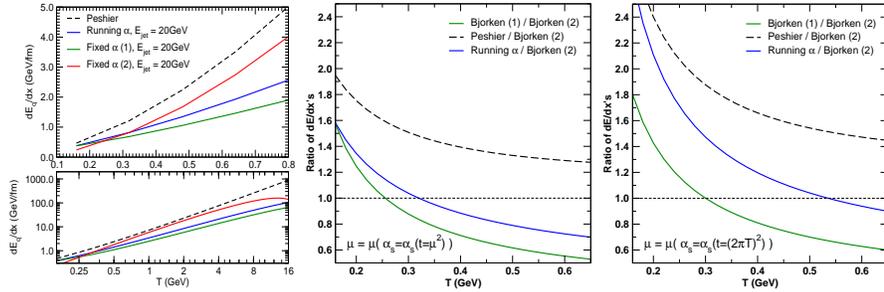
 
\centering
\epsfig{file=dEdxcomp.eps,height=1.5in,width=1.5in,clip=,angle=0} 
\epsfig{file=ratio.eps,height=1.5in,width=1.5in,clip=,angle=0} 
\epsfig{file=ratiooldmu.eps,height=1.5in,width=1.5in,clip=,angle=0} 
\caption{ 
\label{fig:runalpha} 
Left: the average energy loss for a light quark jet evaluated for (1) Peshier's running $\alpha_s=\alpha_s(Q^2)$ but infinite energy jet approximation (2) Finite energy running $\alpha_s$, (3) fixed $\alpha_s$ independent of $Q^2$ but evaluated  at $Q^2=(2\pi T)^2$ and (4) fixed $\alpha_s=0.3$. The functional form of $\alpha_s=\alpha_s(Q^2)$ is taken for vacuum at 1-loop as in~\cite{Peshier:2006hi}. Middle and right: the ratio between the different versions for two different evaluations of the Debye screening scale $\mu$, (1) `self-consistent' $\mu$\cite{Peshier:2006hi} and (2) $\mu$ evaluated at a fixed temperature. `Bjorken (2)' is with all parameters evaluated with $\alpha_s = 0.3$.
\vspace{-0.2in}
}
\end{figure}

\vspace{-0.2in}
In \cite{Wicks:2007am}, the change of the fixed QCD coupling $\alpha_s$ from 0.3 to 0.4 was seen to significantly change the $R_{AA}$ precitions from the WHDG model. There has been much recent work on the effect of a running coupling on the collisional energy loss \cite{Peshier:2006hi,Braun:2006vd} (ie $\alpha_s = \alpha_s(Q^2)$). Here, we revisit the collisional energy loss in a similar manner to \cite{Peshier:2006hi}, looking at a simple Bjorken-style estimate~\cite{Bjorken:1982tu}. Bjorken's estimate for the collisional energy loss is:
\begin{equation}
 \frac{dE_{q,g}}{dx} = \left( \frac{2}{3} \right)^{\pm 1} 2 \pi \left( 1 + \frac{1}{6}n_f \right) \alpha^2 T^2 \ln \left( \frac{4TE}{\mu^2} \right)
\label{eqnPesh1}
\end{equation}
In \cite{Peshier:2006hi}, the running coupling version for very high jet energies is given as:
\begin{equation}
\label{eqnPesh2}
 \frac{dE_{q,g}}{dx} = \left( \frac{2}{3} \right)^{\pm 1} 2 \pi \left( 1 + \frac{1}{6}n_f \right) \frac{\alpha(Q^2=\mu^2)}{b_0} T^2
\end{equation}
although this neglects the finite energy kinematic bound on the jet. Adding in this bound to this calculation gives
\begin{equation}
\label{eqnPesh3}
 \frac{dE_{q,g}}{dx} = \left( \frac{2}{3} \right)^{\pm 1} 2 \pi \left( 1 + \frac{1}{6}n_f \right) \alpha(Q^2=4TE)\alpha(Q^2=\mu^2) T^2 \ln \left( \frac{4TE}{\mu^2} \right)
\end{equation}
which is similar in structure to the original fixed coupling estimate. A numerical comparison of equations \ref{eqnPesh1},\ref{eqnPesh2},\ref{eqnPesh3} is shown in \fig{fig:runalpha}. For reasonable temperatures, $T\sim0.2-0.3$GeV, all results are of a similar order of magnitude. For reasonable energies, no qualitatively new behavior is seen (although, as found in \cite{Peshier:2006hi}, the $E\rightarrow\infty$ behavior is new, but this affects much higher energies than those of interest at RHIC or even LHC). When the kinematic bounds are taken into account, the result for the average energy loss including running coupling is often larger than the fixed $\alpha_s=0.3$ result used in \cite{Wicks:2005gt}. However, the numerical result is very sensitive to the input parameters used, illustrated in the middle and right panes of \fig{fig:runalpha} for changing the prescription for $\mu$ from that used in \cite{Wicks:2005gt} to that from \cite{Peshier:2006hi}.
\vspace{-0.2in}
\vspace{-0.25in}
\section{Conclusion}
\vspace{-0.2in}
It has been argued previously ``that radiative and elastic average energy losses for heavy quarks were in fact comparable over a very wide kinematic range''\cite{Wicks:2005gt}, and even ``E $\geq$ 10 GeV light and charm quark jets have elastic energy losses smaller but of the same order of magnitude as the inelastic losses''\cite{Wicks:2005gt}. Hence, collisional energy loss cannot be neglected when considering jet quenching of high energy jets in the QGP at either RHIC or LHC. A simple model combining collisional and radiative energy losses significantly reduces the discrepancy between the predictions and data.

Two possible improvements to the WHDG model have been examined here. The inclusion of a finite number of collisions is seen to reduce the effect of the collisional energy loss on the quenching of gluons, light and charm quarks, but not to significantly affect the bottom quark $R_{AA}$. Opposite to this effect, including a running QCD coupling increases the energy loss by up to a factor of $~1.5$. The combination of these two affects, along with other large uncertainties in the prediction for electron $R_{AA}$ such as the ratio of charm to bottom total cross-sections, hints at the possibility that both the pion and electron $R_{AA}$s may both be within range of purely perturbative calculations.

\end{document}